\documentclass[letterpaper, 10pt, conference]{ieeeconf}  
\IEEEoverridecommandlockouts                              
\usepackage{amsmath} 
\usepackage{amssymb}  

\usepackage[pdftex]{graphicx}

\usepackage{url}

\def\Ref#1{(\ref{#1})}

\newcommand{\cN}{{\mathcal N}}

\newcommand{\nn}{\nonumber}
\newcommand{\al}[1]{\begin{eqnarray}#1\end{eqnarray}}
\newcommand{\eq}[1]{\begin{equation}#1\end{equation}}

\newcommand{\alequal}{&\!\!\!=\!\!\!&}
\newcommand{\alnequal}{&\!\!\!\neq\!\!\!&}
\newcommand{\alarrow}{&\!\!\!\rightarrow\!\!\!&}
\newcommand{\alequaln}{&\!\!\!\!\!\!\!\!\!=\!\!\!\!\!\!\!\!\!&}

\newcommand{\alnothing}{&\!\!\! \!\!\!&}

\newcommand{\albeginning}{&&\!\!\!\!\!\!\!\!\!\!\!\!\!\!\!\!\!\!\!\!\!}
\newcommand{\smallnegspace}{\!\!\!\!}

\newcommand{\Maha}{Mahalanobis}
\newcommand{\tabwidth}{0.5\columnwidth}

\title{\LARGE \bf
A Complete Derivation Of The Association Log-Likelihood Distance For Multi-Object Tracking
}

\author{Richard Altendorfer
\thanks{R. Altendorfer is with Driver Assistance Systems,
ZF TRW, Koblenz, Germany
        {\tt\small richard.altendorfer@trw.com}}   
 and Sebastian Wirkert
\thanks{S. Wirkert is with German Cancer Research Center, Heidelberg, 
        Germany
        {\tt\small s.wirkert@dkfz.de}}
}

\begin{document}

\maketitle
\thispagestyle{empty}
\pagestyle{empty}

\begin{abstract}

The \Maha\ distance is commonly used in multi-object trackers for measurement-to-track association.
Starting with the original definition of the \Maha\ distance we review its use in association. Given that there is no principle in multi-object tracking that sets the \Maha\ distance apart as a distinguished statistical distance we revisit the global association hypotheses of multiple hypothesis tracking as the most general association setting. Those association hypotheses induce a distance-like quantity for assignment which we refer to as 
association log-likelihood distance.

We compare the ability of the Mahalanobis distance to the association log-likelihood distance to yield correct association relations in Monte-Carlo simulations. It turns out that on average the distance based on association log-likelihood performs better than the Mahalanobis distance, confirming that the maximization of global association hypotheses is a more fundamental approach to association than the minimization of a certain statistical distance measure.

\end{abstract}

\section{INTRODUCTION}

In multi-object trackers where probability distributions of individual tracks are filtered 
new sensor measurements need to be assigned to already existing tracks (or used for object initialization) before the filter
update can be performed. This requires association, i. e. the establishment of relations between filtered, predicted tracks and 
sensor measurements.\footnote{Association is not necessary in approaches where multi-object distributions are filtered 
and tracks are identified and extracted after filtering (see e. g. \cite{Mahler03}).} Despite of the many different aspects and techniques of association such as one-to-one versus one-to-many or 
single-scan versus multi-scan association it always involves the establishment of global association hypotheses and the optimization with respect to hypothesis probabilities. In single-scan association this turns out to be equivalent to computing a measure of distance between tracks and measurements and minimizing
the ``cost" of association by minimization the sum of those distances. In this paper we provide a complete derivation of single-hypothesis assignment from global, joint association events which induces a distance measure for assignment algorithms. We then compare this distance measure with the well-known Mahalanobis distance which is commonly used for measurement-to-track association, see e. g. \cite{Cox_93,stiller2000multisensor,bertozzi2004pedestrian,mahlisch2006heterogeneous}.

Distance measures for measurement-to-track association for multi-object tracking that are variants (depending upon the track life cycle assumptions) of the logarithm of the association likelihood have been presented in \cite{Blackman86}.
In \cite{blanco2012alternative} the association log-likelihood distance (referred to as matching likelihood) was compared to the Mahalanobis
distance in the context of data association for SLAM. The authors' main point is that the association log-likelihood should be used instead of the Mahalanobis distance. While we agree with this statement, we believe that for a proper derivation of the association log-likelihood distance for data association one must start with the
correct expression of the probability for the global joint association event as explained in section \ref{section_derivation_single_hypothesis_assignment}.

The two main contributions of this paper are: 1. a complete derivation of the association log-likelihood distance for measurement-to-track association starting from first principles, and 2. a comparison of association performance using the association log-likelihood distance and the Mahalanobis distance in the context of multi-object tracking.

This paper is organized as follows: first we revisit the origin of the \Maha\ distance and its use in association. We then review global association hypotheses of multiple hypothesis tracking and show how they naturally lead to a distance-like measure for assignment (association log-likelihood distance) which is similar but not identical to the Mahalanobis distance.
By performing Monte-Carlo simulations with both the Mahalanobis distance and the association log-likelihood distance we compare their efficacy in obtaining correct association relations.

\section{The Mahalanobis distance and its role in association}

The Mahalanobis distance was proposed in 1936 \cite{Mahalanobis_36} in connection with hypothesis testing to 
assess the distance\footnote{As in the original paper we use the term distance and not distance squared.} between two probability distributions with a common covariance matrix $\Sigma$ but different means
$\mu_1,\mu_2$
\eq{
d^2_{Maha}\left(\mu_1, \mu_2, \Sigma\right) = \left( \mu_1 - \mu_2 \right)^\top \Sigma^{-1} \left( \mu_1 - \mu_2 \right)
}
It is also used to compute a distance between a sample $s$ (or a sample mean) and a distribution characterized by the mean vector $\mu$ and covariance matrix $\Sigma$
\eq{
d^2_{Maha}\left(s, \mu, \Sigma\right) = \left( s - \mu \right)^\top \Sigma^{-1} \left( s - \mu \right)
}
It is easy to show (see e. g. \cite{bar2004estimation}) that if $x$ is an $n$-dimensional random Gaussian vector with mean $\mu$ and covariance matrix $\Sigma$ then $d^2_{Maha}\left(x, \mu, \Sigma\right)$ is distributed according to a $\chi^2_n$-distribution which enables hypothesis testing with the chi-square test. 

In track-to-track fusion this distance definition is often extended to a distance between
two distributions with different covariance matrices $\left(\mu_1, \Sigma_1\right)$, $\left(\mu_2, \Sigma_2\right)$:
\al{
d^2_{Maha}\left(\mu_1, \mu_2, \Sigma\right) \alequaln \left( \mu_1 - \mu_2 \right)^\top \Sigma^{-1} \left( \mu_1 - \mu_2 \right) \nn\\
\downarrow \alnothing \nn\\
d^{2,gen}_{Maha}\!\!\left(\mu_1, \Sigma_1, \mu_2, \Sigma_2\right) \alequaln \left( \mu_1 - \mu_2 \right)^\top \!\!\left( \Sigma_1 + \Sigma_2 \right)^{-1} \!\left( \mu_1 - \mu_2 \right) \nn
}
in order to provide a distance measure for association between two independent tracks, see e. g. \cite{bar1981track}.

\begin{figure}[ht]
\centering
\includegraphics[viewport = 4cm 8.5cm 18cm 19cm, clip, width = 1 \columnwidth]{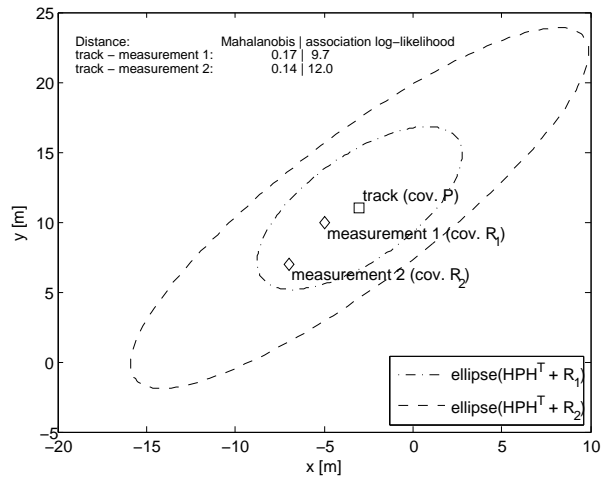}
\caption{``Stealing'' of track: measurement 2 has a large covariance matrix and hence a smaller Mahalanobis distance to the track than measurement 1 despite the fact that the mean of measurement 1 is much closer. Using the Mahalanobis distance would lead to an association between measurement 2 and the track, whereas using the association probability distance would lead to an association between measurement 1 and the track.}
\label{fig_Stealing_Measurements}
\end{figure}

The Mahalanobis distance is commonly used for measurement-to-track association:
the expected measurement $z(k)$ at time $k$ given the predicted state $\xi^-$ and the white, Gaussian measurement noise $r$ is (measurement/output model)
\eq{
z(k) = h(\xi^-(k)) + r(k)
}
Then the probability density function (pdf) of $z$
using linearization is given by $\cN( z; h(\xi^-), H P^- H^\top + R )$ where $H = \partial_\xi h$ is the linearization of the measurement/output function $h(.)$ with respect to the state, and $P^-$ and $R$ are the covariance matrices of $\xi^-$ and $r$, respectively.
Hence the Mahalanobis distance between a measurement sample $z$ and its predicted distribution is
\eq{
d^2_{Maha}\left(z, h(\xi^-), H P^- H^\top + R \right)
}
In nearest-neighbor, single hypothesis association a matrix with distances between all tracks and measurements is computed which is then passed to an assignment algorithm such as the Munkres \cite{munkres1957algorithms} or the auction \cite{bertsekas1988auction} algorithm. 
On the other hand has it been recognized that association using the \Maha\ distance is not optimal since large uncertainties, i. e. large covariance matrices, could lead to very small distances; in particular an uncertain track with large covariance matrix could ``steal" a measurement from another track whose difference of the means is much smaller than the one from the uncertain track, see \cite{Stueker_04,blanco2012alternative}. The complementary situation is depicted in fig. \ref{fig_Stealing_Measurements} where an uncertain measurement (measurement 2) has a smaller Mahalanobis distance to the track than measurement 1 which has a smaller covariance and whose mean is closer to the track. This effect relies on different covariance matrices for different measurements -- while this seems to be unusual for landmark measurements in SLAM \cite{blanco2012alternative} this is common in multi-object tracking.

Therefore, in the next section we will derive the association log-likelihood distance - different from the Mahalanobis distance - used for assignment starting with the probabilities for joint association events as used in multiple hypothesis tracking (MHT). MHT is considered here as the most general case as it encompasses false detections/false tracks as well as new tracks.

\section{Deriving single-hypothesis assignment from joint association events}
\label{section_derivation_single_hypothesis_assignment}

Starting with the joint association probability used for MHT (see \cite{reid1979algorithm,barShalom2011trackingAndDataFusion}) we show how
the individual measurement-track hypotheses (measurement belongs to an existing track, existing track is not detected/observed, measurement belongs to a non-existing track - i. e. a false measurement, measurement belongs to a new track) can be cast into a matrix form suitable for application of single-hypothesis (zero scan) assignment algorithms. This amounts to a pruning of all but the most probable hypothesis at every scan.

The probability of the joint association event $\theta(k)$ (at time $t_k$) for MHT with the number of false measurements and the number of new tracks given by Poisson distributions reads \cite{reid1979algorithm,barShalom2011trackingAndDataFusion}
\al{
P(\theta(k)|Z^k)\alequaln {1 \over c} \prod_{i=1}^{N_{DT}}p\left(z_{j_{i}}(k)|\theta_{j_{i} i}(k), Z^{k-1}\right) P_D\left(\xi^{-}_{i}(k)\right)  \nn\\
\alnothing \cdot\prod_{i=N_{DT}+1}^{N_T}\left( 1 - P_D\left(\xi^{-}_{i}(k)\right) \right) \nn\\
\alnothing \cdot\prod_{j=N_{DT}+1}^{N_{DT}+N_{FD}} \beta_{FD}\left(z_{j}(k)\right)  \nn\\
\alnothing \cdot\prod_{j=N_{DT}+N_{FD}+1}^{N_{D}} \beta_{NT}\left(z_{j}(k)\right)  \label{eq_joint_association_probability}
}
where $c$ is a dimensional normalization constant and the other symbols are explained in table \ref{tab_joint_asso_probability}. Instead of using binary index variables that depend upon $\theta(k)$ as in \cite{reid1979algorithm,barShalom2011trackingAndDataFusion} we assume here that the tracks have been ordered (detected tracks, missed tracks) as well as the detections/measurements (detections of tracks, false detections, new detections). Because of the subsequent restriction to single-hypothesis assignment we have omitted the recursion with respect to the joint cumulative event $\Theta^{k-1,s}$.

\begin{table}[ht]
\caption{Glossary for joint association probability}
\centering
\begin{tabular}{l||l}
\hline
symbol & explanation \\
\hline\hline
$N_{T}$ & \begin{minipage}{\tabwidth}number of established tracks at $t_{k-1}$ \end{minipage} \\
\hline
$N_{DT}$ & \begin{minipage}{\tabwidth}number of established tracks detected at $t_{k}$ \end{minipage} \\
\hline
$N_{D}$ & \begin{minipage}{\tabwidth}number of detections/measurements at $t_{k}$ \end{minipage} \\
\hline
$N_{FD}$ & \begin{minipage}{\tabwidth}number of false detections at $t_{k}$ \end{minipage} \\
\hline
$\theta_{j_i i}(k)$ & \begin{minipage}{\tabwidth} association event that measurement $j_i$ originated from track $i$  \end{minipage} \\
\hline
$P_D\left(\xi^{-}_{i}(k)\right)$ & \begin{minipage}{\tabwidth} probability of detection of track $i$ with predicted state $\xi^{-}_{i}(k)$ \end{minipage} \\
\hline
$\beta_{FD}\left( z_{j}(k) \right)$ & \begin{minipage}{\tabwidth} probability density of a false detection at $z_{j}(k)$ (parameter from Poisson distribution) \end{minipage} \\
\hline
$\beta_{NT}\left( z_{j}(k) \right)$ & \begin{minipage}{\tabwidth} probability density of new track appearance at $z_{j}(k)$ (parameter from Poisson distribution). Here, as in \cite{reid1979algorithm}, the probability of detection $P_D$ has been included in $\beta_{NT}$. The number of new tracks is $N_{NT} = N_{D} - N_{DT} - N_{FD}$.
\end{minipage} \\
\hline
$p\left(z_{j_i}(k)|\theta_{j_i i}(k), Z^{k-1}\right)$ & \begin{minipage}{\tabwidth} association probability density of a measurement $j_i$ given its origin $i$:
$\cN(z_{j_i}; h(\xi_i^-), H^\top P_i^- H + R_{j_i})$, see Appendix \ref{app_conditional_pdf} \end{minipage} \\
\hline
\end{tabular}\label{tab_joint_asso_probability}
\end{table}

By setting the number of new tracks to zero: $N_{NT} = 0\Rightarrow N_{DT}+N_{FT} = N_D$ the expression \Ref{eq_joint_association_probability} reduces to the joint association probability of JPDA \cite{barShalom2011trackingAndDataFusion}. Since we are not interested in maintaining several hypotheses over time but want to pick the most probable joint hypothesis in every scan in a computationally efficient way, we want to cast the individual hypotheses in a form suitable for existing, efficient assignment algorithms that put the global association hypotheses in order and/or find the hypothesis with the maximal probability.
We first apply the logarithm to this joint probability density in order to turn products into sums. However, this requires dimensionless probabilities and not probability densities. Hence we multiply probability densities with an 
arbitrary measurement volume $V_z$. This will not alter the result of the ordering of the hypotheses.
Then the logarithmized probabilities can be cast into a square matrix $a$ as follows\footnote{The last $N_D$ rows are necessary to make the matrix square so that for each of the first $N_T$ columns an independent element must be picked.} (probability arguments have been omitted):
\al{
&&\!\!\!\!\!\!\!\!\!\!\! a = \label{eq_large_assignment_matrix}\\
&& \nn\\
\albeginning \bordermatrix{
              & \substack{{\rm Est.\ tracks}\\ 1\dots N_T}	& \smallnegspace\substack{{\rm False\ tracks}\\ 1\dots N_D} & \smallnegspace\substack{{\rm New\ tracks}\\ 1\dots N_D} \cr
\substack{{\rm Detections}\\ 1\dots N_D\\ {}} & \{\ln(p(\cdot |\cdot) {V_z} P_D)\} & \smallnegspace\{\ln\left(\beta_{FD}{V_z}\right)\} & \smallnegspace\{\ln\left(\beta_{NT}{V_z}\right)\} \cr
\substack{{\rm Missing\ det.}\\ 1\dots N_T\\ {}} & \{\ln\left(1 - P_D\right)\} & \smallnegspace\{0\} & \smallnegspace\{0\} \cr
\substack{{\rm Missing\ det.}\\ 1\dots N_D\\ {}} & \{\ln\left(1 - P_D\right)\} & \smallnegspace\{0\} & \smallnegspace\{0\} \cr
} \nn
}
The set of joint association probability logarithms consists of all $(N_T + 2 N_D)!$ sums of matrix entries that are independent.\footnote{A set of elements of a matrix are said to be independent if no two of them lie in the same row or column \cite{munkres1957algorithms}.}
This is the form required for assignment algorithms such as the Munkres algorithm.
However, this is not an efficient way of representing hypotheses as it leads to many identical hypotheses. For example for one established track and two detections one obtains $(1 + 2\cdot 2)! = 120$ hypotheses of which only eight are distinct, see eq. \Ref{eq_number_of_hypotheses}. By subtracting $\ln\left(1 - P_D\right)$ from the first $N_T$ columns a matrix with zero entries in the last $N_T + N_D$ rows is obtained. 
Now the contribution of independent elements from the last $N_T + N_D$ rows to global hypotheses is zero. 
Truncating the matrix by removing those last $N_T + N_D$ trivial rows and making the non-diagonal entries of $\{\ln\left(\beta_{FD}{V_z}\right)\}$ and $\{\ln\left(\beta_{NT}{V_z}\right)\}$ forbidden leads to a smaller, rectangular $N_D\times(N_T + 2 N_D)$ matrix with
\eq{
\sum_{N_{DT}=0}^{\min(N_{D}, N_T)} \sum_{N_{FD}=0}^{N_D - N_{DT}}\binom{N_D}{N_{DT}}\binom{N_D - N_{DT}}{N_{FD}}{ {N_{T}}! \over (N_{T} - N_{DT})!}  \label{eq_number_of_hypotheses}
}
distinct hypotheses.
The resulting hypotheses are identical to eq. \Ref{eq_joint_association_probability} except for a common factor which is irrelevant for the ordering of the hypotheses.

For the remainder of this paper we focus on the original matrix \Ref{eq_large_assignment_matrix}.
The matrix entries are the logarithmized association probabilities between tracks (existing, false, new) and measurements (existing, non-existing). In particular, the logarithmized association probability between an existing track $i$ and an existing measurement $j$ reads
\al{
a_{ji} \alequal \ln\left( \cN(z_{j}; h(\xi_i^-), H P_i^- H^\top + R_{j}) {V_z} \right) + \ln\left( P_D \right) \nn\\
\alequal -{1\over 2}{\Delta z_{ji}}^\top \Sigma_{\Delta z_{ji}}^{-1}{\Delta z_{ji}} 
 + \ln\left( V_z \over \sqrt{|2\pi \Sigma_{\Delta z_{ji}}|} \right) \nn\\
\alnothing + \ln\left( P_D \right)\nn
}
where $\Delta z_{ji} = z_{j} - h(\xi_i^-)$ and $\Sigma_{\Delta z_{ji}} = H P_i^- H^\top + R_{j}$.
Multiplying the matrix \Ref{eq_large_assignment_matrix} by $-2$ turns maximization of the sum of independent matrix entries into minimization; choosing $V_z = 1\cdot units(V_z)$ yields the following expression
\al{
d^2_{{Asso-LL,ji}} \alequal -2 a_{ji} \nn\\
       \alequal \underbrace{{\Delta z_{ji}}^\top \Sigma_{\Delta z_{ji}}^{-1}{\Delta z_{ji}}}_{=d^2_{Maha}} + \ln{\left|\Sigma_{\Delta z_{ji}}\right|\over units\left(V^2_z\right) } \nn\\
			\alnothing + n \ln\left( 2 \pi \right) - 2 \ln\left( P_D \right)  \label{eq_inno_based_distance}
}
whose first term is the Mahalanobis distance and where $n=\dim\left( \Delta z \right)$. We refer to $d^2_{Asso-LL}$ as the association log-likelihood distance because $p\left(z_{j_i}(k)|\theta_{j_i i}(k), Z^{k-1}\right)$ is conditioned on the association event $\theta_{j_i i}$. Setting $P_D = 1$ gives the logarithm of the probability density of the predicted measurement. Note that in settings where sensor models with different dimensions $n$ are used at the same time the term $n \ln( 2 \pi )$ will have an influence on global association hypotheses.

\begin{figure}[ht]
\centering
\includegraphics[viewport = 4cm 8.5cm 18cm 19cm, clip, width = 1 \columnwidth]{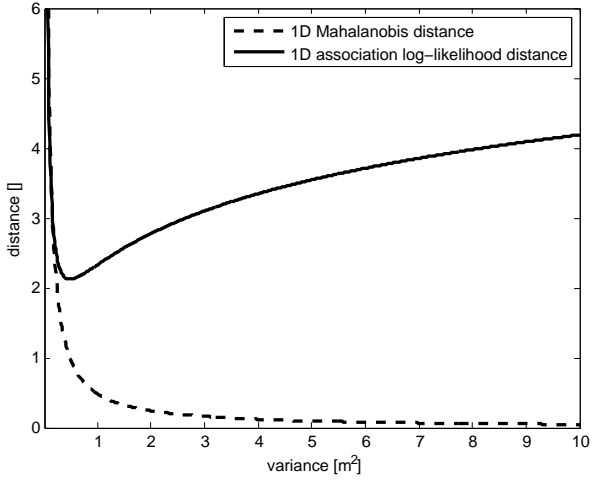}
\caption{One-dimensional generalized Mahalanobis and association log-likelihood distances as a function of the sum of variances, with $\Delta z = 0.7 m$.}
\label{fig_Maha_and_innovation}
\end{figure}

Using this association probability distance tracks or measurements with large covariances are penalized in association with respect to tracks with small ones, see fig. \ref{fig_Maha_and_innovation}, thereby avoiding the situation depicted in fig. \ref{fig_Stealing_Measurements} where a rather uncertain track with large covariance has a smaller Mahalanobis distance to the measurement than a second track whose mean is much closer to the measurement. 

The association distance \Ref{eq_inno_based_distance} can hence be derived from first principles -- in particular the term $\ln{\left|\Sigma_{\Delta z_{ji}}\right|}$ need not be viewed as a heuristic distance penalty \cite{Stueker_04,Maehlisch_et_al_08}.

{-- \it Relation to results presented in \cite{blanco2012alternative}: }
As mentioned in the introduction the case for the association log-likelihood distance compared to the Mahalanobis distance has been made in \cite{blanco2012alternative}. However, their expression for the probability of the joint association event is different from the one used here (eq. \Ref{eq_joint_association_probability}, see also \cite{reid1979algorithm,barShalom2011trackingAndDataFusion}). In particular note that $P(\theta(k)|Z^k) = P(\theta(k)|Z(k), Z^{k-1})$ depends upon the history of all current and past measurements.
This is {\it not} equal to 
\eq{
P(\theta(k)|Z(k), \{\xi^{-}(k)\}) \label{eq_joint_asso_prob_tracks}
}
where $\{\xi^{-}(k)\}$ denotes the set of $N_T$ established, predicted tracks at time $k$. The difference is that by conditioning on tracks instead of measurements the track estimation uncertainty subsumed in its covariance matrix is missing.
Indeed, using the probability \Ref{eq_joint_asso_prob_tracks} the factors of the likelihood function of $\theta(k)$ on the right-hand side of eq. \Ref{eq_joint_association_probability} change:
\al{
\albeginning p\left(z_{j_{i}}(k)|\theta_{j_{i} i}(k), Z^{k-1}\right) \nn\\
\alarrow p\left(z_{j_{i}}(k)|\theta_{j_{i} i}(k), \{\xi^{-}(k)\}\right) \nn\\
\alequal p\left(z_{j_{i}}(k)|\xi_i^{-}(k)\right) \nn\\
\alequal \cN(z_{j_i}(k); H\xi_i^-(k), R_{j_i}) \nn\\
\alnequal \cN(z_{j_i}(k); H\xi_i^-(k), H P_i^- H^\top + R_{j_i}) \nn
}
Therefore expression \Ref{eq_joint_asso_prob_tracks} as well as the analogous expression in \cite[eq. 1]{blanco2012alternative} should not be the starting point for deriving an association distance.\footnote{However, the authors 
arrive at the correct result by equating $p\left(z_{j_{i}}(k)|\theta_{j_{i} i}(k), \{\xi^{-}(k)\}\right)$ to $\cN(z_{j_i}(k); H\xi_i^-(k), H P_i^- H^\top + R_{j_i})$ in \cite[eq. 2]{blanco2012alternative} instead of $\cN(z_{j_i}(k); H\xi_i^-(k), R_{j_i})$. Here, we have translated their expressions into the analogous ones in our notation.}

\section{Comparison of distances}

Despite evidence for the association log-likelihood distance detailed above the \Maha\ distance has also been commonly used for association (see e. g. \cite{Cox_93,stiller2000multisensor,bertozzi2004pedestrian,mahlisch2006heterogeneous}). This motivates our investigation how the association log-likelihood distance compares with the Mahalanobis distance with respect to their performance in association: 
\al{
d^2_{Maha}\left(s, \mu, \Sigma\right) \alequal \left( s - \mu \right)^\top \Sigma^{-1} \left( s - \mu \right) \nn\\
d^2_{Asso-LL}\left(s, \mu, \Sigma\right) \alequal \left( s - \mu \right)^\top \Sigma^{-1} \left( s - \mu \right) \nn\\
         \alnothing + \ln{\left|\Sigma \right|\over units\left(\left|\Sigma\right|\right)} \nn\\
				 \alnothing + n \ln\left( 2 \pi \right) - 2 \ln\left( P_D \right) \label{eq_d_Asso_LL}
}

We want to compare the efficacy of the distances listed above in making correct measurement-to-track associations. We focus on a single-scan, single time-step set-up. For each time-step we create $N$ ground truth tracks of the form $\xi_i, i=1, ..., N$ (see \Ref{eq_state_vector}) in a certain state space volume using a uniform distribution. Then we draw $N$ samples for current measurements from $\cN(z_i; h(\xi_i), R_i)$ and $N$ current estimated predicted states from $\cN(\hat\xi_i^-; \xi_i, P_i^-)$ where the expressions for the covariance matrices $R_i$ and $P_i^-$ are obtained as detailed below.
The optimal Munkres association algorithm is used to find the best association between these measurements and predicted tracks using the distance matrix created by the respective generalized distance.
This set-up enables us to cleanly separate the influence of the used association distance on assignment from the specifics of tracking and track management, in particular we do not simulate tracks over time in their life cycle stages (initialized, under test, established, ...) and set $P_D = 1$.

The performance of the association algorithms is evaluated at three levels of track separation, i. e. number of tracks $N\in \{10,30,50\}$ in a given state volume, and with two different measurement models specified in eq. \Ref{eq_measurement_models}. Also, in order to incorporate the effect of the track life cycle stages on the state covariance we investigate two extreme cases:

\paragraph{State covariance matrix from steady state}
For every ground truth track $i$ we generate its estimated predicted covariance matrix $P^-_{i,\infty}$ at steady state by using the dynamical system specified in Appendix \ref{app_vehicleModel} where the process noise covariance matrix $V_i$ and the measurement noise covariance matrix $R_i$ are used as input for the discrete algebraic Riccati equation:
\al{
P^-_{i,\infty} \alequal F P^-_{i,\infty} F^\top \nn\\
\alnothing - F P^-_{i,\infty} H^\top \left( H P^-_{i,\infty} H^\top + R_i\right)^{-1} H P^-_{i,\infty} F^\top \nn\\
\alnothing + G V_i G^\top
}
Then, the estimated track state mean $\hat\xi_i^-$ is obtained as a sample of $\cN(\hat\xi_i^-; \xi_i, P^-_{i,\infty})$. The measurement $z_i$ is obtained as a sample of $\cN(z_i; h(\xi_i), R_i)$.
The covariance matrices $V_i$ and $R_i$ are constructed by a diagonal matrix with uniformly distributed positive definite entries which is then rotated with a uniformly distributed angle in $[0, 2\pi]$.

\paragraph{State covariance matrix with arbitrary shape}
While a multi-measurement, multi-track scenario at steady state is at one end of the spectrum of possible life cycle stages we also investigate a case where the predicted state covariance matrix is not at steady state but can assume an arbitrary shape. In this case, we construct $P^-_{i}$ as $V_i$ and $R_i$ above.

The ground truth tracks were drawn from a state volume of $V_\xi = [-20 m, 20 m]\times [-20 m, 20 m]\times [-40 m/s, 40 m/s]\times [-40 m/s, 40 m/s]$.
The simulation was run in ten batches with 10000 association scenarios each; over those ten batches the deviation in the resulting percentage of correct assignments from the mean was less than $0.4\%$.

\begin{figure}[ht]
\centering
\includegraphics[viewport = 4cm 8.5cm 18cm 19cm, clip, width = 1 \columnwidth]{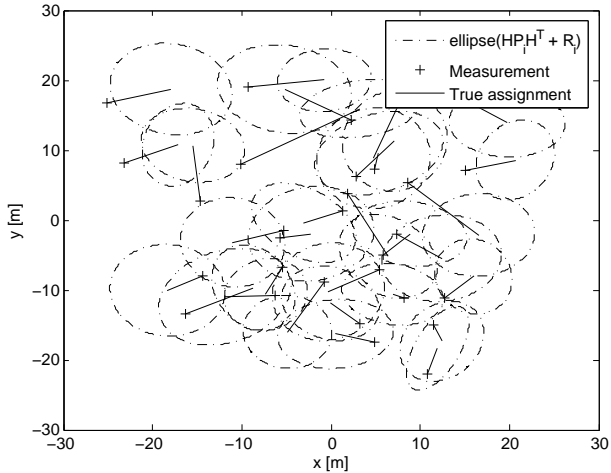}
\caption{Error ellipses of measurements and predicted tracks of a sample simulation with $N=30$ tracks.}
\label{fig_Association_simulation}
\end{figure}

The simulation results can be found  in table \ref{tab_results_steady_state_H1H2} for the predicted tracks being in a steady state and in table \ref{tab_results_arbitrary_H1H2} for the predicted tracks with arbitrary covariance shape. In fig. \ref{fig_Association_simulation} a simulation sample with $N=30$ tracks is shown.

Note that the proposed set-up assumes one global association hypothesis with the largest probability is picked -- we do not maintain several hypotheses as in MHT.  The creation of measurements for already existing tracks also means that we are restricting ourselves to the upper, left sub-matrix of eq. \Ref{eq_large_assignment_matrix} with $N_D = N_T$ and $P_D = 1$.

\begin{table}[ht]
\caption{Correct assignment rate [\%] for tracks in steady state}
\centering
\begin{tabular}{|c|c|c|c|c|c|c|}
\hline
Number of tracks & \multicolumn{2}{c|}{$N=10$} & \multicolumn{2}{c|}{$N=30$} & \multicolumn{2}{c|}{$N=50$} \\
Distance\textbackslash Output $H$& $H_1$ & $H_2$ & $H_1$ & $H_2$ & $H_1$ & $H_2$ \\
\hline\hline
$d^2_{Maha}$                      &  79.3 & 79.8 & 49.8 & 50.9 & 34.5 & 35.6 \\
\hline
$d^2_{Asso-LL}$ &  81.9 & 82.3 & 55.0 & 56.0 & 40.5 & 41.5 \\
\hline
\end{tabular}
\label{tab_results_steady_state_H1H2}
\end{table}

For tracks in steady state the association log-likelihood distance has a significant advantage with respect to the Mahalanobis distance.
This advantage becomes larger (in a relative sense) as the number of tracks increases. The choice of measurement/output model $H_i$ has little effect on the assignment rate (table \ref{tab_results_steady_state_H1H2}).

\begin{table}[ht]
\caption{Correct assignment rate [\%] for tracks with arbitrary covariance shape}
\centering
\begin{tabular}{|c|c|c|c|c|c|c|}
\hline
Number of tracks & \multicolumn{2}{c|}{$N=10$} & \multicolumn{2}{c|}{$N=30$} & \multicolumn{2}{c|}{$N=50$} \\
Distance\textbackslash Output $H$& $H_1$ & $H_2$ & $H_1$ & $H_2$ & $H_1$ & $H_2$ \\
\hline\hline
$d^2_{Maha}$                      &  72.3 & 70.8 & 39.2 & 37.6 & 25.9 & 24.7 \\
\hline
$d^2_{Asso-LL}$       &  72.4 & 70.8 & 39.4 & 37.8 & 26.2 & 24.9 \\
\hline
\end{tabular}
\label{tab_results_arbitrary_H1H2}
\end{table}

For tracks with an arbitrary covariance shape we do not observe significant differences between the association log-likelihood and the Mahalanobis distance. Again the choice of measurement/output model $H_i$ has little effect on the assignment rate (table \ref{tab_results_arbitrary_H1H2}).

Next, we want to explore the influence of the term proportional to the measurement dimension $n$ in $d^2_{Asso-LL}$ \Ref{eq_d_Asso_LL}. For tracks in steady state using $H_1$ as measurement model we consider a scenario in which for some measurement-track pairs $(z_j, \xi^-_i)$, namely with $j$ and $i$ both odd, association is attempted with an only one-dimensional measurement defined by the first row of $H_1$: $H_{11}$. For all other index combinations the full two-dimensional measurement with $H_1$ is used. In order to assess the influence of $n \ln( 2 \pi )$ we perform association with $d^2_{Maha}$ and $d^2_{Asso-LL}$ as well as with the association log-likelihood distance without the $n \ln( 2 \pi )$-term: $d^2_{Asso-LL} - n \ln( 2 \pi )$.

\begin{table}[ht]
\caption{Correct assignment rate [\%] for tracks in steady state with one- and two-dimensional measurements}
\centering
\begin{tabular}{|c|c|c|c|}
\hline
Number of tracks                  & $N=10$ & $N=30$ & $N=50$ \\
Distance\textbackslash Output $H$ & $H_1,H_{11}$ & $H_1,H_{11}$ & $H_1,H_{11}$ \\
\hline\hline
$d^2_{Maha}$                      &  72.1 & 40.2 & 27.4  \\
\hline
$d^2_{Asso-LL}$       						&  79.8 & 53.4 & 40.9  \\
\hline
$d^2_{Asso-LL} - n \ln( 2 \pi )$  &  79.0 & 51.7 & 39.2  \\
\hline
\end{tabular}
\label{tab_results_steady_state_different_dimensions}
\end{table}

In table \ref{tab_results_steady_state_different_dimensions} we observe that the omission of the term proportional to the measurement dimension $n \ln( 2 \pi )$ does have a detrimental effect on the correct assignment rate. Also in this scenario with mixed measurement models the comparative advantage of $d^2_{Asso-LL}$ with respect to $d^2_{Maha}$ becomes larger compared to table \ref{tab_results_steady_state_H1H2}.

\begin{table}[ht]
\caption{Correct assignment rate [\%] for tracks with arbitrary covariance shape with one- and two-dimensional measurements}
\centering
\begin{tabular}{|c|c|c|c|}
\hline
Number of tracks                  & $N=10$ & $N=30$ & $N=50$ \\
Distance\textbackslash Output $H$ & $H_1,H_{11}$ & $H_1,H_{11}$ & $H_1,H_{11}$ \\
\hline\hline
$d^2_{Maha}$                      &  65.7 &  32.6 &  21.4  \\
\hline
$d^2_{Asso-LL}$       						&  73.2 &  42.3 &  29.4  \\
\hline
$d^2_{Asso-LL} - n \ln( 2 \pi )$  &  72.2 &  40.7 &  28.2  \\
\hline
\end{tabular}
\label{tab_results_arbitrary_shape_different_dimensions}
\end{table}

Again, the omission of the term proportional to the measurement dimension $n \ln( 2 \pi )$ has a detrimental effect on the correct assignment rate (table \ref{tab_results_arbitrary_shape_different_dimensions}). Also, in contrast to the results in table \ref{tab_results_arbitrary_H1H2}, we observe a significantly better assignment rate using $d^2_{Asso-LL}$ as compared to $d^2_{Maha}$.

\section{Conclusions}

Starting with the probability density of the joint association event for MHT we have derived the association log-likelihood distance to be used in assignment algorithms. Given the fact that the Mahalanobis distance is also commonly used for assignment we have investigated both distances in terms of their association performance
by performing Monte-Carlo simulations. In steady-state scenarios the association log-likelihood distance performed significantly better than the Mahalanobis distance. In scenarios with predicted track covariance matrices of arbitrary shape the association log-likelihood distance exhibited a better behavior when different measurement dimensions were present; when all measurements were two-dimensional the comparative advantage of the association log-likelihood distance over the Mahalanobis distance was within the statistical fluctuation of $<0.4\%$.

In summary in this paper we have pointed out that the Mahalanobis distance has no special role in assignment algorithms and we have shown that the association log-likelihood distance is the one that can be derived from first principles in multi-object tracking.
The association log-likelihood distance also performs significantly better in steady state scenarios. 
This supports our proposition that maximization of global association hypotheses is a more fundamental approach to association than the minimization of a certain statistical distance measure.






\begin{appendix} 
\label{app_dynamical_system}

\subsection{Dynamical system}
\label{app_vehicleModel}

The vehicle kinematics is characterized by a four-dimensional state vector
\eq{
\xi = \begin{pmatrix} x & y & \dot x & \dot y \end{pmatrix}^\top \label{eq_state_vector}
}
The dynamical model is a discrete-time counterpart white noise acceleration model \cite{LiJilkov03}
\eq{
\xi^-(k+1) = F(\Delta t_k) \xi(k) + G( \Delta t_k )\nu(k) \label{eq:dynamicalModel}
}
where
\al{
F(\Delta t_k) &=& \begin{pmatrix}  1 & 0 & \Delta t_k & 0 \cr
                        0 & 1 & 0 & \Delta t_k  \cr
                        0 & 0 & 1 & 0 \cr
                        0 & 0 & 0 & 1 \end{pmatrix} \nn\\
G(\Delta t_k) &=& \begin{pmatrix} \frac{\Delta t_k^2}{2} & 0\cr
                        0 & \frac{\Delta t_k^2}{2}\cr
                        \Delta t_k & 0\cr
                        0 & \Delta t_k \end{pmatrix}
}
with $\Delta t_k = t_{k+1} - t_k$.
The process noise $\nu(k)$ is modeled by a white, mean-free Gaussian process with covariance matrix $V(k)$.

The sensor is modeled by a 2-D position measurement with
\eq{
z(k) = h(\xi^-(k)) + r(k) \label{eq_measurement_equation}
}
where $z(k)$ is the measurement
and the measurement noise $r(k)$ is modeled by a white, mean-free Gaussian process with covariance matrix $R(k)$.
The measurement/output function $h$ is given by a linear function $h(\xi) = H\xi$ where we investigate two different models
\al{
H_1 \alequal \begin{pmatrix} 1 &  0 & 0 & 0 \cr
                             0 &  1 & 0 & 0 \end{pmatrix} \nn\\
H_2 \alequal \begin{pmatrix} 1 & -1 & 0 & 0 \cr
                             0 &  1 & 0 & 0 \end{pmatrix} \label{eq_measurement_models}
}
The measurement function is chosen to be linear in order to solve the Riccati equation for state-independent matrices.

\subsection{Conditional probability density of individual measurement} \label{app_conditional_pdf}

The conditional probability of the measurement $z_{j_i}(k)$ given the set of previous measurements $Z^{k-1}$ and the event $\theta_{j_i i}(k)$ as it appears in eq. \Ref{eq_joint_association_probability} can be computed as follows for a linear system using Gaussian expressions for $p\left(z_{j_i}(k)|\theta_{j_i i}(k), \xi_i \right)$ and $p\left(\xi_i|Z^{k-1}\right)$ and using the Markov assumption:
\al{
&&\!\!\!\!\!\!\!\!\!\!\!\! p\left(z_{j_i}(k)|\theta_{j_i i}(k), Z^{k-1}\right) \nn\\
\alequaln \int_{\xi_i} p\left(z_{j_i}(k)|\theta_{j_i i}(k), Z^{k-1}, \xi_i \right) p\left(\xi_i|\theta_{j_i i}(k), Z^{k-1}\right) d\xi_i \nn\\
\alequaln \int_{\xi_i} p\left(z_{j_i}(k)|\theta_{j_i i}(k), \xi_i \right) p\left(\xi_i|Z^{k-1}\right) d\xi_i \nn\\
\alequaln \int_{\xi_i} \cN(z_{j_i}(k); H\xi_i, R_{j_i}) \cN(\xi_i; \hat\xi^{-}_i(k), P_i^-) d\xi_i  \nn\\
\alequaln \cN(z_{j_i}(k); H\hat\xi_i^-(k), H P_i^- H^\top + R_{j_i})
}
Alternatively, this as well as the expression for a nonlinear measurement function $h$:
$\cN(z_{j_i}(k); h(\hat\xi_i^-(k)), H P_i^- H^\top + R_{j_i})$ can be derived by propagating the probability densities for $\hat\xi^-$ and $r$ through eq. \ref{eq_measurement_equation}.

\end{appendix}


\bibliographystyle{IEEEtran}
\bibliography{bibliography}

\end{document}